\documentclass[conference]{IEEEtran}
\IEEEoverridecommandlockouts
\usepackage{cite}
\usepackage{amsmath,amsthm,amssymb,amsfonts}
\usepackage{algorithmic}
\usepackage{graphicx}
\usepackage{textcomp}
\usepackage{xcolor}
\usepackage{listings}
\usepackage{hyperref}
\usepackage{braket}
\usepackage{subcaption}

\def\BibTeX{{\rm B\kern-.05em{\sc i\kern-.025em b}\kern-.08em
    T\kern-.1667em\lower.7ex\hbox{E}\kern-.125emX}}

\begin{document}

\title{Block-encodings as programming abstractions: \newline The Eclipse Qrisp \texttt{BlockEncoding} Interface\\}

\author{
\IEEEauthorblockN{Matic Petri\v{c}}
\IEEEauthorblockA{
%\textit{dept. name of organization (of Aff.)} \\
\textit{\textit{Fraunhofer Institut FOKUS}}\\
Berlin, Germany \\
matic.petric@fokus.fraunhofer.de}
\and
\IEEEauthorblockN{René Zander}
\IEEEauthorblockA{
%\textit{dept. name of organization (of Aff.)} \\
\textit{Fraunhofer Institut FOKUS}\\
Berlin, Germany \\
rene.zander@fokus.fraunhofer.de}
}

\maketitle

\begin{abstract}
Block-encoding is a foundational technique in modern quantum algorithms, enabling the implementation of non-unitary operations by embedding them into larger unitary matrices. While theoretically powerful and essential for advanced protocols like Quantum Singular Value Transformation (QSVT) and Quantum Signal Processing (QSP), the generation of compilable implementations of block-encodings poses a formidable challenge. This work presents the \texttt{BlockEncoding} interface within the Eclipse Qrisp framework, establishing block-encodings as a high-level programming abstraction accessible to a broad scientific audience. Serving as both a technical framework introduction and a hands-on tutorial, this paper explicitly details key underlying concepts abstracted away by the interface, such as block-encoding construction and qubitization, and their practical integration into methods like the Childs-Kothari-Somma (CKS) algorithm. We outline the interface's software architecture, encompassing constructors, core utilities, arithmetic composition, and algorithmic applications such as matrix inversion, polynomial filtering, and Hamiltonian simulation. Through code examples, we demonstrate how this interface simplifies both the practical realization of advanced quantum algorithms and their associated resource estimation.
\end{abstract}


\section{Introduction}

Quantum computers natively execute operations that are strictly reversible and unitary. However, many practical applications require the application of non-unitary operators to quantum states. Block-encoding resolves this fundamental limitation by embedding a subnormalized non-unitary operator into a larger-dimensional unitary matrix. This technique has become a cornerstone for a wide array of advanced quantum algorithms. For instance, it is instrumental in the quantum linear systems solver zoo, ranging from the foundational HHL algorithm \cite{harrow2009quantum}, the CKS algorithm \cite{childs2017quantum}, to QSVT \cite{gslw_qsvt_19}, Zeno eigenstate filtering \cite{lin2020optimal}, and the near-optimal Dalzell linear solver \cite{dalzell2024shortcut}, as thoroughly detailed in recent survey \cite{morales2024quantum}. Furthermore, recent advances in Quantum Oracle Sketching promise exponential quantum advantages for processing massive classical data \cite{zhao2026exponential}. Beyond linear systems, block-encodings are essential for simulating open quantum systems \cite{an2023linear} and applying polynomial filter transformations \cite{lee2025filtered, filip2026beyond}.

Despite its theoretical elegance and its foundational role in the Grand Unification of Quantum Algorithms \cite{mrtc_unification_21}, the practical realization of block-encodings remains a non-trivial software engineering challenge. Developers must manually manage auxiliary (ancilla) qubits, calculate subnormalization factors, and synthesize complex multi-qubit control logic. Furthermore, while quantum signal processing applications leverage established libraries such as QSPPACK \cite{qsppack} and pyqsp \cite{pyqsp} for calculating optimal phase angles, developers face an additional bottleneck: these mathematical outputs must then be manually integrated into a separate quantum software framework to synthesize the final executable circuits.

Recognizing these integration bottlenecks, prominent quantum software ecosystems such as Qualtran \cite{harrigan2024qualtran} and PennyLane \cite{bergholm2018pennylane}, alongside emerging domain-specific languages like Cobble \cite{yuan2025cobble}, have introduced dedicated abstractions to streamline block-encoding generation. Contributing to this effort toward accessible, high-level quantum programming, we introduce block-encoding as a programming abstraction within the Eclipse Qrisp framework \cite{seidel2024qrisp} through the \texttt{BlockEncoding} class, enabling compilation and execution on physical quantum hardware.
%To bridge this gap, we introduce block-encodings as a high-level programming abstraction within the Eclipse Qrisp framework \cite{seidel2024qrisp} through the \texttt{BlockEncoding} class. 
This paper details the architecture and practical usage of this interface. The remainder of this work is organized as follows: Section~\ref{sec:preliminaries} provides the mathematical background of block-encodings, qubitization, and block encoding Chebyshev polynomials. Section~\ref{sec:blockencoding} details the programming abstraction and the underlying software interface, encompassing instantiation, utilities, and arithmetic composition. Section~\ref{sec:algorithms} discusses the built-in algorithmic applications, while Section~\ref{sec:examples} evaluates the framework's practical utility and resource estimation capabilities through concrete code examples.

\section{Mathematical necessities}
\label{sec:preliminaries}

\subsection{Block-encodings}

A block-encoding of a (not necessarily unitary) operator $\hat{A}$ acting on a system Hilbert space $\mathcal{H}_s$ is a unitary operator $\hat{U}_A$ acting on an extended Hilbert space $\mathcal{H}_a \otimes \mathcal{H}_s$, where $\mathcal{H}_a$ represents an auxiliary (ancilla) variable.

Conceptually, the operator $\hat{A}$ is embedded in the upper-left block of $\hat{U}_A$ such that:
\begin{equation}
\hat{U}_A = 
\begin{pmatrix} 
 \hat{A}/\alpha & * \\ 
 * & *
\end{pmatrix}
\end{equation}

More formally, $\hat{U}_A$ is called an $(\alpha, m, \epsilon)$-block-encoding of $\hat{A}$, if \cite{lin2022lecturenotesquantumalgorithms}:
\begin{equation}
\|\hat{A} - \alpha (\bra{0}_a \otimes \hat{\mathbb{I}}_s) \hat{U}_A (\ket{0}_a \otimes \hat{\mathbb{I}}_s) \| \leq \epsilon
\end{equation}
where $\alpha \geq \|\hat{A}\|$ is the subnormalization factor ensuring the singular values of $\hat{A}/\alpha$ fall within the unit disk, and $\epsilon \ge 0$ is the approximation error, and $m$ are the ancillary qubits.

Consequently, applying the operator $\hat{A}$ to a quantum state $\ket{\psi}_s$ involves preparing the joint system in $\ket{0}_a \otimes \ket{\psi}_s$, applying $\hat{U}_A$, and post-selecting the ancilla variable on the zero-state $\ket{0}_a$. The probability of successfully measuring the ancilla in the zero-state is $P_{\text{success}} = \|\hat{A}\ket{\psi}\|^2 / \alpha^2$.

While we assume projection onto the standard zero-state, this framework can equivalently be defined involving arbitrary auxiliary state preparations, often denoted as a unitary-operator pair $(\hat U, \hat G)$ such that $(\bra{G}_{a} \otimes \hat{\mathbb{I}}_{s}) \hat{U} (\ket{G}_{a} \otimes \hat{\mathbb{I}}_{s}) = \hat{H}$ where $\ket{G}_a=\hat{G}\ket{0}_a$ \cite{low2019hamiltonian,kirby2023exact}.

In this notation, constructing a block-encoding via linear combination of unitaries (LCU) is described as follows: Let $\hat{A}=\sum_k\alpha_k\hat{U}_k$, where $\alpha_k>0$ are positive coefficients and $\hat{U}_k$ are unitaries. Define the unitaries
\begin{equation*}
    \hat{G}=\sum_k\sqrt{\alpha_k}\ket{k}_a\bra{0}_a,\quad \hat{U}=\sum_k\ket{k}_a\bra{k}_a\otimes U_k.
\end{equation*}
The unitary $\hat{U}$ acts as a quantum switch case (SELECT) applying the $k$-th unitary $U_k$ conditioned on the ancialla variable being in state $\ket{k}_a$, and $\hat{G}$ prepares (PREP) the state $\ket{G}=\hat{G}\ket{0}_a$.

\begin{figure}[htbp]
    \centering
    \includegraphics[width=\columnwidth]{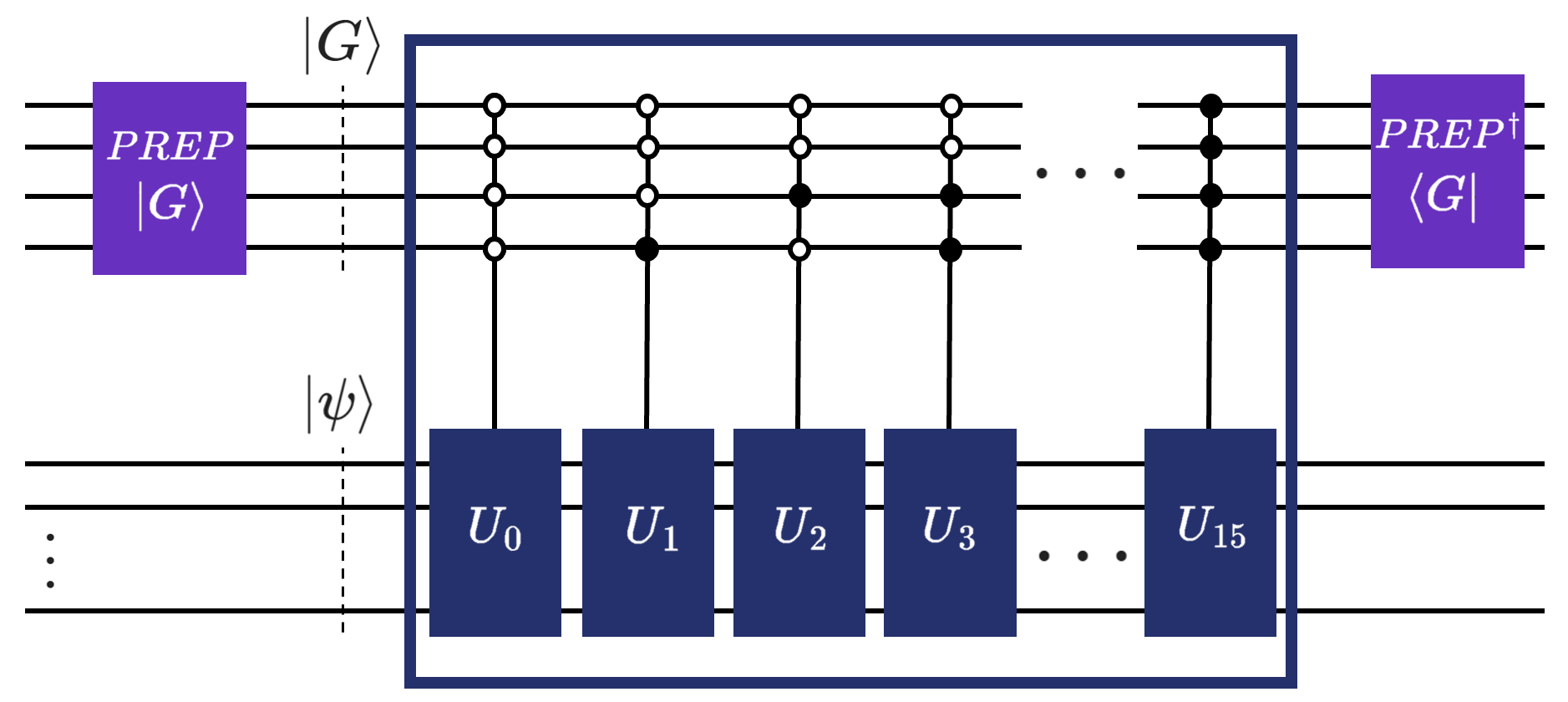}
    \caption{Visual schematics of the block-encoding construction via LCU. Important to note that Qrisp's \texttt{q\_switch} utilizes a more efficient $SELECT$ procedure using balanced binary trees \cite{Khattar2025riseofconditionally}.}
    \label{fig:block_encoding_lcu}
\end{figure}

\subsection{Qubitization and Chebyshev polynomials}

While a block-encoding provides a ``static'' representation of an operator $\hat{A}$, qubitization is a technique that transforms this encoding into a unitary quantum walk operator $\hat{\mathcal W}$. For a Hermitian operator $\hat{A}$, this walk operator encodes the eigenvalues of the normalized operator $\hat{A}/\alpha$ into its rotational phases, enabling transformations of the spectrum through Quantum Signal Processing \cite{low2019hamiltonian}.

Given an exact block-encoding $(\hat{U}, \hat{G})$ of a Hermitian operator $\hat{A}$ such that $\hat{U}^2 = \hat{\mathbb{I}}$ (i.e., the block-encoding unitary $\hat{U}$ is Hermitian), we define a reflection operator $\hat{R}$ acting on the auxiliary space as:
\begin{equation}
\hat{R} = (2|G\rangle_a\langle G|_a - \hat{\mathbb{I}}_a) \otimes \hat{\mathbb{I}}_s
\end{equation}
The qubitized walk operator $\hat{\mathcal{W}}$ is then constructed by interleaving the block-encoding unitary with this reflection:
\begin{equation}
\hat{\mathcal{W}} = \hat{R} \hat{U}
\end{equation}

Rigorous spectral analysis\cite{low2019hamiltonian, lin2022lecturenotesquantumalgorithms} demonstrates that for any eigenstate $\ket{\psi_{\lambda}}$ of the normalized operator $\hat{A}/\alpha$ with eigenvalue $\lambda \in [-1, 1]$, the walk operator $\hat{\mathcal{W}}$ acts invariantly on a two-dimensional subspace (hence the term ``qubitization'') spanned by $\ket{G}_a \otimes \ket{\psi_{\lambda}}_s$ and an orthogonal state. Within this invariant subspace, the eigenvalues of $\hat{\mathcal{W}}$ are:
\begin{equation}
\mu_\pm = \lambda \pm i\sqrt{1-\lambda^2} = e^{\pm i \arccos(\lambda)}
\end{equation}
By lifting the eigenvalues onto the unit circle, qubitization allows the application of polynomial transformations to the spectrum of $\hat{A}$ via repeated applications of $\hat{\mathcal{W}}$ \cite{low2019hamiltonian}.

\begin{figure}[htbp]
    \centering
    \includegraphics[width=\columnwidth]{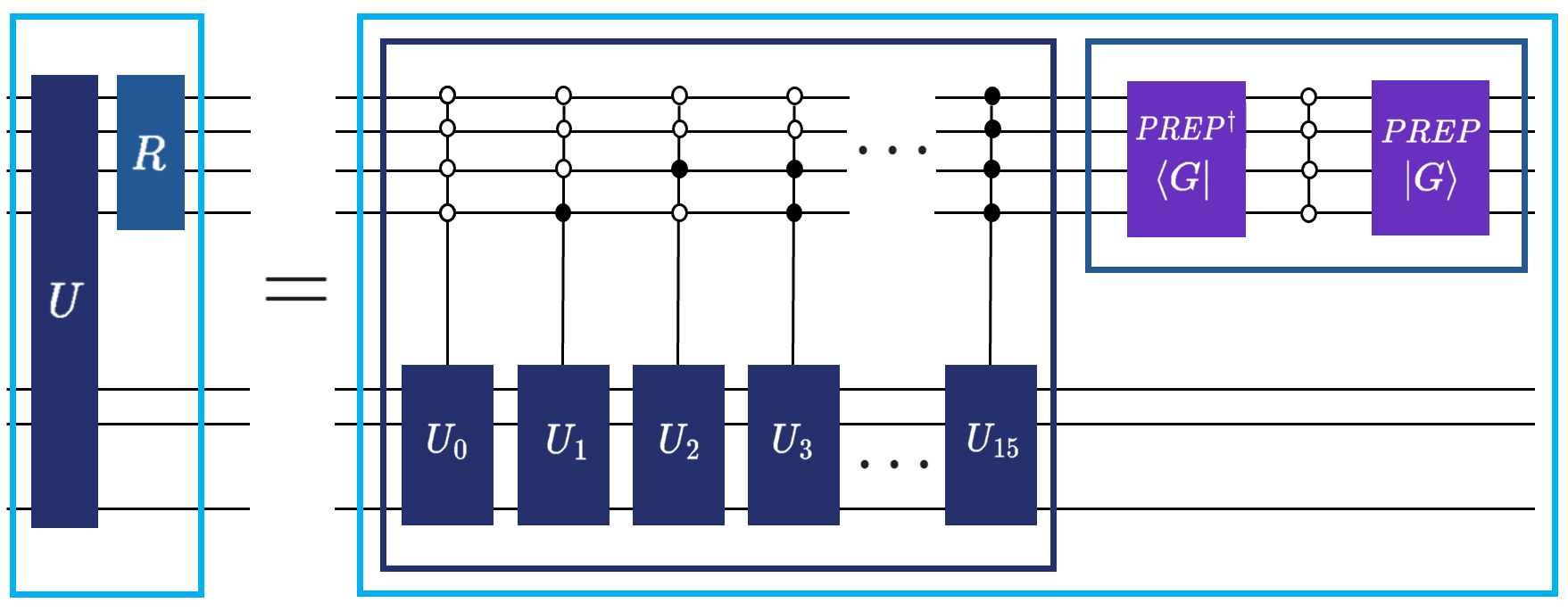}
    \caption{Visual schematics of one application of the qubitization walk operator $\hat{\mathcal{W}}$ as a sequence of $\hat U$ followed by a reflection operator $\hat{R}$.}
    \label{fig:qubitization}
\end{figure}

One of the most powerful applications of the qubitization framework is its connection to Chebyshev polynomials of the first kind, $T_k(x)$. Because the eigenvalues of the walk operator $\hat{\mathcal{W}}$ are $e^{\pm i \arccos(\lambda)}$, iteratively applying $\hat{\mathcal{W}}$ exactly $k$ times, induces a phase of $e^{\pm i k \arccos(\lambda)}$. Projecting this back onto the signal state $|G\rangle_a$ extracts the real part of this transformation, corresponding to $T_k(\hat{A}/\alpha)$. Formally:
\begin{equation}
(\langle G|_a \otimes \hat{\mathbb{I}}_s) \hat{\mathcal{W}}^k (|G\rangle_a \otimes \hat{\mathbb{I}}_s) = T_k(\hat{A}/\alpha)
\end{equation}
for any integer $k \ge 0$ \cite{low2019hamiltonian}.
This construction is highly advantageous for quantum algorithmic design: Evaluating $T_k(\hat{A}/\alpha)$ requires exactly $k$ applications of the underlying block-encoding unitary $\hat{U}$. Furthermore, because Chebyshev polynomials provide the optimal uniform approximation to continuous functions over the domain $[-1, 1]$, this direct algebraic access to $T_k$ guarantees near-optimal resource scaling for advanced algorithms such as the quantum Lanczos method \cite{kirby2023exact}.

\begin{figure}[htbp]
    \centering
    \includegraphics[width=\columnwidth]{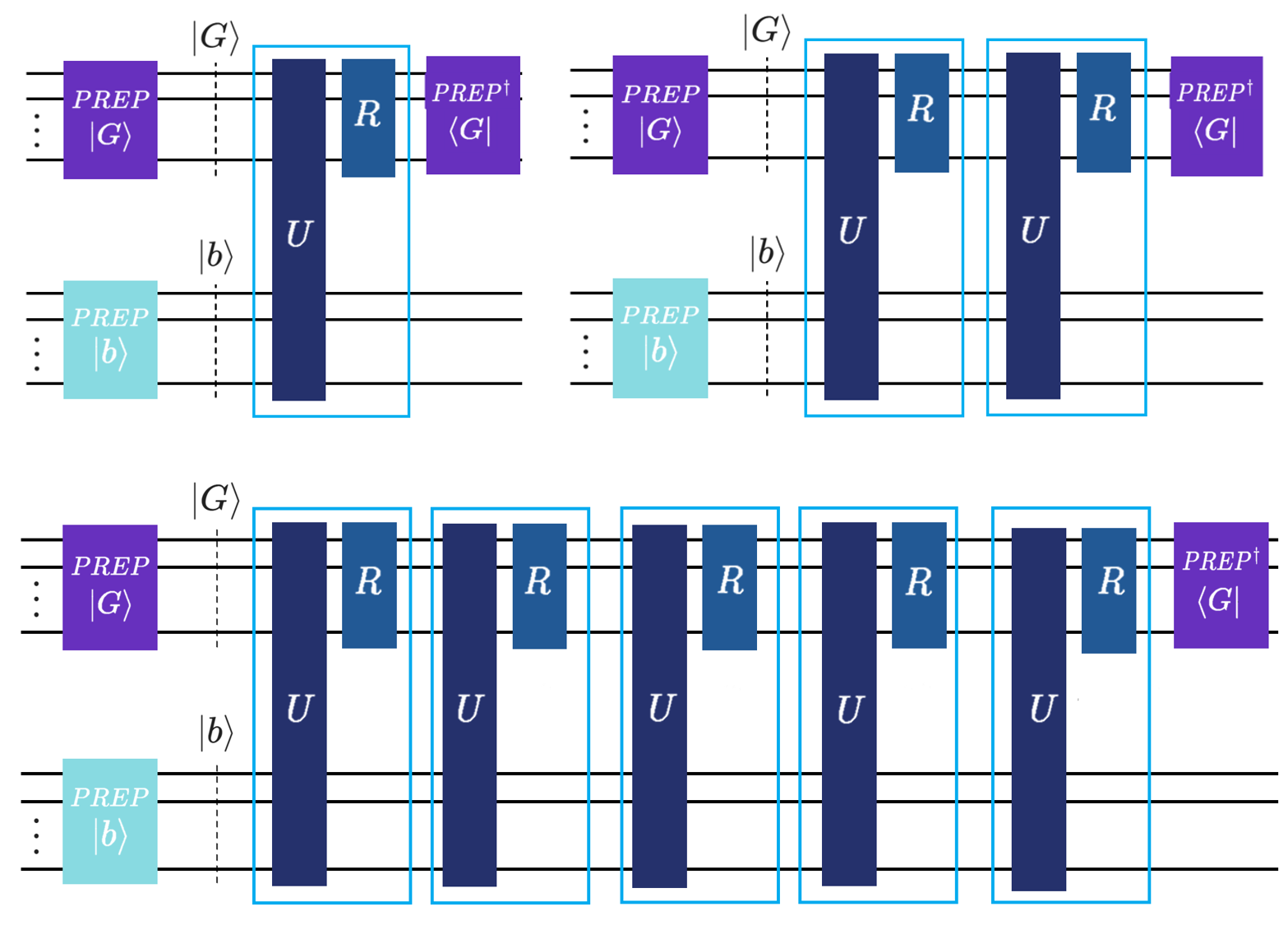}
    \caption{Visual schematics of block-encoding the $k$-th Chebyshev polynomial of the first kind $T_k$ through repeated application of the qubitization walk operator $\hat{\mathcal{W}}^k$ for $k=1, 2, 5$.}
    \label{fig:block_encoding_cheb}
\end{figure}

\section{The Qrisp BlockEncoding Interface}
\label{sec:blockencoding}

For a detailed introduction to Qrisp's abstractions allowing to program via \texttt{QuantumVariables} and functions rather than direct quantum circuit manipulation, we refer the reader to \cite{seidel2024qrisp}.

The \texttt{BlockEncoding} class in Qrisp abstracts the mathematical formalisms described in Section~\ref{sec:preliminaries} into a seamless object-oriented interface. The core attributes include the scaling factor \texttt{alpha} which, the \texttt{unitary} function representing $\hat{U}_A$, and a boolean flag \texttt{is\_hermitian} indicated whether the block-encoding unitary is Hermitian.

\subsection{Constructors}

The \texttt{BlockEncoding} class provides several factory methods to instantiate block-encodings from various mathematical structures:

\begin{itemize}
    \item \texttt{from\_array(A)}: Constructs a block-encoding directly from a 2-D numerical matrix $\hat{A}$.

    \item \texttt{from\_operator(O)}: Constructs a block-encoding directly form quantum mechanical operators, i.e., Qrisp's \texttt{QubitOperator} and \texttt{FermionicOperator}.
    
    \item \texttt{from\_lcu(coeffs, unitaries, \dots)}: Utilizes the linear combination of unitaries protocol to encode $\hat{A} = \sum_k \alpha_k \hat{U}_k$.
    
    \item \texttt{from\_eye(k)}\footnote{At the time of publication, this feature is available in the GQSVT branch of the Eclipse Qrisp repository: \url{https://github.com/eclipse-qrisp/Qrisp/tree/GQSVT}. \label{gqsvt_branch}}: Constructs a block-encoding of a 2-D matrix with ones on the diagonal and zeros elsewhere. The integer \texttt{k} refers to the index of the diagonal. \texttt{k=0} (default) refers to the main diagonal (identity operator $\hat{\mathbb{I}}$), a positive value refers to an upper diagonal, and a negative number to a lower diagonal.

    \item \texttt{from\_projector(left, right, \dots)}\footref{gqsvt_branch}: Constructs a block-encoding of a projector $\ket{\phi}\bra{\psi}$, where the arguments \texttt{left} and \texttt{right} are either integers indicating computational basis states, or functions preparing the states $\ket{\phi}$ and $\ket{\psi}$, respectively. It can also construct the kernel projector $\hat{\mathbb{I}} - \ket{\phi}\bra{\phi}$.
    
    \item \texttt{BlockEncoding(alpha, anciallas, unitary, \dots)}: Initializes a custom block-encoding by explicitly defining the subnormalization factor $\alpha$, ancilla \texttt{QuantumVariable} templates, and the \texttt{unitary} function.
\end{itemize}

\subsection{Core Utilities}

Once instantiated, the \texttt{BlockEncoding} interface provides methods for application and quantum resource estimation:

\begin{itemize}
    \item \texttt{.apply(operands)}: Applies the block-encoding unitary $\hat{U}_A$ to a given set of operand \texttt{QuantumVariables} in state $\ket{\psi}$, automatically allocating and managing the required ancilla variables. This application is probabilistic and is only successful if these ancillas are subsequently measured (post-selected) in the $\ket{0}$ state.
    
    \item \texttt{.apply\_rus(operand\_prep)}: Implements a repeat-until-success loop by repeatedly preparing the initial state and applying the encoded operator until the ancilla variables are measured in the $\ket{0}$ state. If the measurement yields any other result, the run is discarded and the process restarts. The \texttt{operand\_prep} argument is a function that initializes the \texttt{QuantumVariables} in state $\ket{\psi}$ for each attempt.

    \item \texttt{.resources(operands, \dots)}: Performs resource estimation for the block-encoding circuit without requiring full state-vector simulation. This method returns hardware metrics, i.e., gate counts, circuit depth, and maximum qubit allocation.

    \item \texttt{.expectation\_value(operand\_prep, \dots)}: Measures the expectation value $\langle \psi | \hat{A} | \psi \rangle$ using the Hadamard test protocol. It applies the function \texttt{operand\_prep} to initialize the operand \texttt{QuantumVariables} in state $\ket{\psi}$.
    
    \item \texttt{.qubitization()}: Returns a \texttt{BlockEncoding} representing the qubitization walk operator $\hat{\mathcal{W}}$, which embeds the spectrum of $\hat{A}$ as phases in the eigenvalues of a unitary.
    
    \item \texttt{.chebyshev(k, \dots)}: Returns a \texttt{BlockEncoding} representing the $k$-th Chebyshev polynomial of the first kind applied to the operator, i.e., $T_k(\hat{A})$.
\end{itemize}

\subsection{Arithmetic Composition}

Another feature of the Qrisp \texttt{BlockEncoding} interface is its overloading of Python's arithmetic operators, enabling intuitive algebraic composition of block-encodings. If $\hat{A}$ and $\hat{B}$ are block-encodings, the framework automatically handles the underlying quantum circuit synthesis to create new block-encodings for:

\begin{itemize}
    \item \textbf{Addition/Subtraction} (\texttt{A + B} / \texttt{A - B}): Employs linear combination of unitaries (LCU) techniques to encode $\hat{A} \pm \hat{B}$.
    These operations require both block-encodings to have the same operand structure (e.g., number, size of variables).
    
    \item \textbf{Scalar Multiplication} (\texttt{A * scalar}): Encodes $c\hat{A}$ by updating the subnormalization factor $\alpha \to |c|\cdot\alpha$. If the scalar $c$ is negative, a phase shift of $\pi$ is applied.
    
    \item \textbf{Negation} (\texttt{-A}): Implements the transformation $\hat{A} \to -\hat{A}$ by applying a phase shift of $\pi$.
    % to the state while maintaining the original subnormalization factor $\alpha$.
    
    \item \textbf{Matrix Multiplication} (\texttt{A @ B}): Encodes the operator product $\hat{A}\hat{B}$. The resulting subnormalization factor is the product of the individual factors ($\alpha_{new} = \alpha_A \alpha_B$).
    
    \item \textbf{Tensor Product} (\texttt{A.kron(B)}): Encodes the Kronecker product $\hat{A} \otimes \hat{B}$. The new block-encoding acts jointly on the combined operand variables for both $\hat{A}$ and $\hat{B}$.
\end{itemize}

The inner workings of these arithmetic methods are currently implemented following the approaches from \cite{dalzell2023quantum}, with the integration of the more efficient product decompositions from \cite{dong2025products} planned for future releases.
Given a single block-encoded operator $\hat{A}$, these primitives could theoretically be applied to construct a block-encoding of a polynomial $p(\hat{A})$. However, doing so incurs significant resource overhead. Consequently, these arithmetic operations should be used sparingly, primarily for multivariate compositions (e.g., forming $\hat{A}+\hat{B}$). Once composed, we encourage users to utilize the \texttt{poly(coeffs, kind)} method to evaluate complex polynomials of the combined/single operator, achieving more efficient encodings via Generalized Quantum Signal Processing \cite{motlagh2024generalized, sunderhauf2023generalized}. Future updates will fully automate this workflow, compiling all algebraic compositions into the most resource-efficient encoding by default.

\section{Algorithmic Applications}
\label{sec:algorithms}

The \texttt{BlockEncoding} interface serves as a functional foundation for high-level quantum algorithms, providing out-of-the-box methods for complex mathematical transformations of a block-encoded operator $\hat{A}$:

\begin{itemize}
    \item \texttt{.poly(coeffs, kind)}: Encodes the operator $p(\hat{A})$ for an arbitrary complex polynomial $p(x)$. The \texttt{kind} keyword argument allows the polynomial $p(x)$ to be defined using either standard monomial or Chebyshev bases.

    \item \texttt{.inv(eps, kappa)}: Encodes an approximation of the inverse operator $\hat{A}^{-1}$ with precision $\epsilon$, given an upper bound on the condition number $\kappa$.
    
    \item \texttt{.sim(t, N)}: Encodes an approximation of time-evolution operator $e^{-it\hat{A}}$ for a given simulation time $t$ and truncation order $N$.
\end{itemize}

\subsection{Polynomial Transformations}
Through the \texttt{.poly()} method (Listing~\ref{lst:poly}), users can apply arbitrary polynomial functions $p(x)$ to the eigenvalues of a Hermitian encoded operator $\hat{A}$ using generalized quantum eigenvalue transformation (GQET) \cite{sunderhauf2023generalized}. For specific spectral manipulations, \texttt{.chebyshev()} applies the $k$-th Chebyshev polynomial of the first kind $T_k(\hat{A})$. Both methods rely internally on the \texttt{.qubitization()} walk operator $\hat{\mathcal{W}}$ to achieve these spectral transformations.

\subsection{Matrix Inversion}
The \texttt{.inv()} method (Listing~\ref{lst:inv}) approximates the inverse $\hat{A}^{-1}$ of an encoded Hermitian operator utilizing quantum eigenvalue transformation (QET) to implement a polynomial approximation of $f(x) = 1/x$ over the domain $D_{\kappa} = [-1, -1/\kappa] \cup [1/\kappa, 1]$. This method provides an abstraction for solving linear systems of equations ($\hat{A}x = b$). The query complexity scales as $\mathcal{O}(\kappa^2 \log(\kappa/\epsilon))$, where $\epsilon$ is the target precision and $\kappa$ is an upper bound on the condition number \cite{mrtc_unification_21}. To be precise, we assume that all eigenvalues of the rescaled operator $\hat{A}/\alpha$ lie within the domain $D_{\kappa}$.

\subsection{Hamiltonian Simulation}
Simulating the time evolution of a quantum system governed by a Hamiltonian $\hat{H}$ requires implementing the operator $e^{-it\hat{H}}$. The \texttt{.sim()} method (Listing~\ref{lst:sim})  approximates this evolution using the Jacobi-Anger expansion into Bessel functions of the first kind $J_n(t)$, with truncation order $N$. This expansion achieves super-exponential convergence in the truncation order, providing favorable scaling in both simulation time and error compared to standard Trotterization methods, particularly for large evolution times or high-precision requirements \cite{low2019hamiltonian, motlagh2024generalized}.

\smallskip

All aforementioned methods automatically handle the calculation of the quantum signal processing angles based on state-of-the-art techniques \cite{laneve2025generalized, ni2025inverse}. Additionally, Qrisp provides further quantum algorithms built on the \texttt{BlockEncoding} interface; notable examples, illustrated in Section~\ref{sec:examples}, include quantum Lanczos algorithm \texttt{lanczos\_alg}, alongside the \texttt{CKS} and \texttt{dalzell\_inversion} linear system solvers.

\section{Implementation Examples}
\label{sec:examples}

\subsection{NumPy-like syntax}
The Qrisp \texttt{BlockEncoding} interface provides a high-level programming abstraction that translates classical matrix operations for non-unitary operators directly into quantum programs. By overloading standard algebraic operators, the framework allows researchers to work within the familiar language of linear algebra rather than navigating the intricacies of superposition and entanglement.

To demonstrate this modularity, consider the evaluation of a composite matrix expression involving a polynomial transformation and a matrix inversion:
\begin{equation}
    \hat{C} = \hat{\mathbb{I}} + \hat{A} - 2\hat{A}^2 + \hat{B}^{-1}
\end{equation}
In Qrisp, this operator is implemented by first applying the \texttt{.poly()} and \texttt{.inv()} methods to the respective block-encodings $\hat{A}$ and $\hat{B}$, and subsequently combining them via the addition operation. The univariate polynomial $\hat{\mathbb{I}}+\hat{A}-2\hat{A}^2$ could also be implemented utilizing the arithmetic operations, albeit the \texttt{.poly()} method is more efficient. The framework automatically scales the subnormalization constants and manages ancilla variables.

\begin{lstlisting}[caption={NumPy-like syntax for combining block-encodings}, label={lst:numpy}]
import numpy as np
from qrisp import *
from qrisp.block_encodings import BlockEncoding

# 1. Define matrices and condition number
A = np.array([[0.66, 0.02], [0.02, 0.82]])
B = np.array([[0.78, -0.01], [-0.01, 0.57]])
kappa = np.linalg.cond(B)

# 2. Block-encode and evaluate the expression: C = I + A - 2A^2 + B^-1
BA = BlockEncoding.from_array(A)
BB = BlockEncoding.from_array(B)

# .poly([1, 1, -2]) represents the polynomial 1 + 1*x - 2*x^2
BC = BA.poly([1., 1., -2.]) + BB.inv(0.01, kappa)

# Use arithmetic operations instead of poly()
#BI = BlockEncoding.from_eye()
#BC = BI + BA - 2 * BA @ BA + BB.inv(0.01, kappa)

# 3. Apply to a state |b> using Repeat-Until-Success (RUS)
def prep_b():
    qv = QuantumFloat(1)
    prepare(qv, [1, 2]) # State proportional to [1, 2]
    return qv

# Uses Qrisp's state-vector simulator and samples from the output distribution
@terminal_sampling
def main():
    # Use Repeat-Until-Success (RUS) to apply the non-unitary operator C to the state |b>
    return BC.apply_rus(prep_b)()

# 4. Result Comparison
res_dict = main()
amps = np.sqrt([res_dict.get(i, 0) for i in range(2)])
print(amps)
# [0.41719948 0.90881494]
\end{lstlisting}

The algebraic operations (\texttt{+}, \texttt{-}, \texttt{@}) are useful for combining block-encodings of distinct operators $\hat{A}$ and $\hat{B}$.

\begin{lstlisting}[caption={Arithmetic Operations on block-encodings}, label={lst:arithmetic}]
import numpy as np
from qrisp.block_encodings import BlockEncoding

# Define two simple 2x2 matrices
A = np.array([[1, 0], [0, -1]]) # Pauli Z
B = np.array([[0, 1], [1, 0]])  # Pauli X

# Construct initial block-encodings
BA = BlockEncoding.from_array(A)
BB = BlockEncoding.from_array(B)

# 1. Addition and Subtraction (LCU-based)
B_add = BA + BB
B_sub = BA - BB

# 2. Scalar Multiplication and Negation
B_scaled = 2.5 * BA
B_neg = -BA

# 3. Matrix Multiplication
B_mult = BA @ BB

# 4. Tensor Product (Kronecker Product)
B_kron = BA.kron(BB)
\end{lstlisting}

\subsection{Algorithmic applications}

\subsubsection{\texttt{.poly(coeffs, kind)}}
The following example (Listing~\ref{lst:poly}) illustrates the application of the polynomial transformation discussed previously. By directly passing the target coefficients to the method, the framework automatically constructs the necessary GQET circuit to apply $p(A) = 1 + 2A + A^2$ to the block-encoded Hermitian matrix.
A prominent application of such arbitrary polynomial transformations is eigenstate filtering \cite{Zander_filtering,lee2025filtered,filip2026beyond}.

\begin{lstlisting}[caption={Polynomial Transformation using \texttt{.poly}}, label={lst:poly}]
import numpy as np
from qrisp import *
from qrisp.block_encodings import BlockEncoding

# Define a Hermitian matrix A and a vector b
A = np.array([[0.73, 0.14, -0.15, -0.04],
              [0.14, 0.68, -0.05, -0.01],
              [-0.15, -0.05, 0.77, -0.03],
              [-0.04, -0.01, -0.03, 0.59]])
b = np.array([0, 1, 1, 1])

# Generate a block-encoding and apply the polynomial p(A) = 1 + 2A + A^2
BA = BlockEncoding.from_array(A)
BA_poly = BA.poly(np.array([1., 2., 1.]))

def prep_b():
    operand = QuantumVariable(2)
    prepare(operand, b)
    return operand

@terminal_sampling
def main():
    # Use Repeat-Until-Success (RUS) to apply the non-unitary operator p(A) to the state |b>
    return BA_poly.apply_rus(prep_b)()

res_dict = main()
amps = np.sqrt([res_dict.get(i, 0) for i in range(len(b))])
print(amps)
# [0.03835136 0.57233673 0.62852841 0.52527314]
\end{lstlisting}

\subsubsection{\texttt{.inv(eps, kappa)}}
To solve a linear system of equations, developers can utilize the \texttt{.inv()} method as shown in Listing~\ref{lst:inv}. By supplying the target precision ($\epsilon$) and condition number ($\kappa$), the framework generates the inverted block-encoding $\hat{A}^{-1}$. This operator can then be seamlessly applied to a prepared target state $\ket{b}$ using standard algorithmic primitives like the Repeat-Until-Success (RUS) loop.

\begin{lstlisting}[caption={Quantum Linear System Solver using \texttt{.inv}}, label={lst:inv}]
# Assuming A, b, and prep_b are defined as in the previous example
kappa = np.linalg.cond(A)
BA = BlockEncoding.from_array(A)

# Approximate A^-1 with target precision 0.01 and condition number kappa
BA_inv = BA.inv(eps=0.01, kappa=kappa)

@terminal_sampling
def main():
    return BA_inv.apply_rus(prep_b)()

res_dict = main()
amps = np.sqrt([res_dict.get(i, 0) for i in range(len(b))])
print(amps)
# [0.03356433 0.56309959 0.52736387 0.63535788]
\end{lstlisting}

\subsubsection{\texttt{.sim(t, N)}}
To simulate the dynamics of a quantum system, developers can utilize the \texttt{.sim()} method as shown in Listing~\ref{lst:sim}. After generating a block-encoding from a physical observable, such as an Ising Hamiltonian, the user simply supplies the simulation time ($t$) and expansion order ($N$). The framework then generates (an approximation to) the unitary evolution operator $e^{-it\hat{H}}$, which can be directly applied to prepare system's time-evolved state.

\begin{lstlisting}[caption={Hamiltonian Simulation using \texttt{.sim}}, label={lst:sim}]
from qrisp import *
from qrisp.block_encodings import BlockEncoding
from qrisp.operators import X, Z

def create_ising_hamiltonian(L, J, B):
    return sum(-J * Z(i) * Z(i + 1) for i in range(L-1)) + \
           sum(B * X(i) for i in range(L))

L = 4
H = create_ising_hamiltonian(L, 0.25, 0.5)
BE = BlockEncoding.from_operator(H)

@terminal_sampling
def main(t):
    # Evolve state |0> for time t using order N=8
    BE_sim = BE.sim(t=t, N=8)
    return BE_sim.apply_rus(lambda: QuantumFloat(L))()

res_dict = main(0.5)
amps = np.sqrt([res_dict.get(i, 0) for i in range(2**L)])
\end{lstlisting}

\subsection{Quantum Resource Estimation: Discrete Laplace Operator}
We compare a generic block-encoding against a customized Linear Combination of Unitaries (LCU) approach. The target observable is the one-dimensional discrete Laplace operator with periodic boundary conditions. Mathematically, this operator can be efficiently decomposed into $\hat{L} = -2I + \hat{V} + \hat{V}^{\dagger}$, where $\hat{V}\colon\ket{k} = \ket{(k+1) \bmod N}$ and $\hat{V}^{\dagger}\ket{k} = \ket{(k-1) \bmod N}$ represent cyclic adders. While the \texttt{from\_array()} method treats $\hat{L}$ as a generic matrix, the custom implementation utilizes the \texttt{from\_lcu()} constructor to realize these cyclic permutations via efficient arithmetic circuits \cite{gidney2018halving}. As demonstrated in the following Listing~\ref{lst:qre_laplace}, evaluating the circuit complexities via the \texttt{.resources()} method reveals a drastic reduction in both gate count and circuit depth, underscoring the absolute necessity of structure-aware block-encodings for scalable quantum algorithms.

\begin{lstlisting}[caption={Resource Estimation: Comparing Block-Encodings of the Discrete Laplace Operator}, label={lst:qre_laplace}]
import numpy as np
from qrisp import *
from qrisp.block_encodings import BlockEncoding

N = 256
I = np.eye(N)
A = -2*I + np.eye(N, k=1) + np.eye(N, k=-1)
A[0, N-1] = 1
A[N-1, 0] = 1

# Generic block-encoding
BA = BlockEncoding.from_array(A)

# Custom block-encoding
def I(qv):
    gphase(np.pi, qv[0])

def V(qv):
    qv += 1

def V_dg(qv):
    qv -= 1

unitaries = [I, V, V_dg]
coeffs = np.array([2.0, 1.0, 1.0])
BA_custom = BlockEncoding.from_lcu(coeffs, unitaries)

res = BA.resources(QuantumFloat(8))
res_custom = BA_custom.resources(QuantumFloat(8))
print("Generic block-encoding resources:", res)
print("Custom block-encoding resources:", res_custom)
# Generic block-encoding resources: {'gate counts': {'p': 100, 'gphase': 2, 'u3': 510, 'cy': 608, 'cx': 3172, 'h': 572, 'x': 195, 't': 858, 't_dg': 1144}, 'depth': 4789, 'qubits': 24}
# Custom block-encoding resources: {'gate counts': {'p': 1, 'gphase': 2, 'u3': 6, 'cx': 144, 'h': 46, 'x': 5, 's': 14, 't': 34, 't_dg': 36, 'measure': 14}, 'depth': 109, 'qubits': 19}
\end{lstlisting}

\subsection{Quantum Lanczos Method}

The quantum Lanczos method \cite{kirby2023exact} is an algorithm designed to estimate the eigenvalues of a Hamiltonian $\hat{H}$ by constructing a Krylov subspace $\mathcal{K}_k(\hat{H}, \ket{\psi_0}) = \text{span}\{\ket{\psi_0}, \hat{H}\ket{\psi_0} \dots, \hat{H}^{k-1}\ket{\psi_0}$ \cite{kirby2023exact}. In the Qrisp framework, this is facilitated by the \texttt{lanczos\_alg} function. This high-level routine leverages qubitization \cite{low2019hamiltonian} to prepare states within the Krylov subspace $\ket{\psi_k} = T_k(\hat{H})\ket{\psi_0}$. These states are measured to estimate the subspace Hamiltonian entries, which are then diagonalized classically to find the ground state energy.

\begin{lstlisting}[caption={Quantum Lanczos: Solving the Heisenberg Model}, label={lst:lanczos}]
import networkx as nx
from qrisp import QuantumVariable
from qrisp.algorithms.lanczos import lanczos_alg
from qrisp.operators import X, Y, Z
from qrisp.vqe.problems.heisenberg import create_heisenberg_init_function
from qrisp.jasp import jaspify

# 1. Define a 1D Heisenberg model on 6 qubits
L = 6
G = nx.cycle_graph(L)
# H = sum (X_i X_j + Y_i Y_j + 0.5 Z_i Z_j)
H = (1/4)*sum((X(i)*X(j) + Y(i)*Y(j) + 0.5*Z(i)*Z(j)) for i,j in G.edges())

# 2. Prepare initial state function (tensor product of singlets)
M = nx.maximal_matching(G)
U_singlet = create_heisenberg_init_function(M)

def operand_prep():
    qv = QuantumVariable(L)
    U_singlet(qv)
    return qv

# 3. Run the quantum Lanczos algorithm
# D is the Krylov space dimension
D = 6

# Use jaspify for high-performance JIT compilation and tracing
@jaspify(terminal_sampling=True)
def main():
    # lanczos_alg automates block-encoding and diagonalization
    return lanczos_alg(H, D, operand_prep)

energy = main()

# 4. Result Verification
print(f"Ground state energy estimate: {energy}")
print(f"Exact Ground state energy: {H.ground_state_energy()}")
# Ground state energy estimate: -2.3656442004580995                               
# Exact Ground state energy: -2.3680339887499047
\end{lstlisting}

\subsection{Comparing Quantum Linear Solvers}

These algorithms represent the state-of-the-art in Quantum Linear System Solvers (QLSS) \cite{morales2024quantum}, optimizing for target precision $\epsilon$ and condition number $\kappa$. The theoretical scaling of the approaches implemented within Qrisp is summarized in Table~\ref{tab:qlss_scaling}. 

While both the native \texttt{.inv()} method (discussed previously) and the CKS algorithm \cite{childs2017quantum} utilize a Chebyshev polynomial approximation of $f(x) = 1/x$, implemented via QET/QSVT\footref{gqsvt_branch} and LCU, respectively, the Dalzell solver \cite{dalzell2024shortcut} adopts a structurally distinct kernel reflection approach to achieve optimal resource scaling.

\begin{table}[htbp]
\centering
\renewcommand{\arraystretch}{1.15}
\setlength{\tabcolsep}{4pt}
\begin{tabular}{p{0.48\columnwidth} p{0.44\columnwidth}}
\hline
\textbf{Approach} & \textbf{Scaling} \\
\hline
Classical (Conjugate Gradient)\cite{shewchuk1994cg} & $O(N s \kappa \log(1/\epsilon))$ \\
HHL\cite{harrow2009quantum} & $O(\kappa^{2}/\epsilon)$ \\
CKS (LCU with Chebyshev)\cite{childs2017quantum} & $O(\kappa^2\, \log(\kappa/\epsilon))$ \\
QET/QSVT\cite{mrtc_unification_21} & $O(\kappa^2\, \log(\kappa/\epsilon))$ \\
Optimal (Dalzell) solver\cite{dalzell2024shortcut} & $O(\kappa\, \log(1/\epsilon))$ \\
\hline
\end{tabular}
\caption{Theoretical query complexity of quantum linear system solvers implemented in Qrisp. The access models assume a Hamiltonian evolution oracle for HHL and a block-encoding oracle for the polynomial-based solvers. For QET/QSVT and CKS, the query complexity for a single repetition is $\mathcal{O}(\kappa\, \log(\kappa/\epsilon))$, corresponding to the degree of the polynomial transformation. While guaranteeing a successful projection requires $\mathcal{O}(\kappa)$ applications due to post-selection, the CKS algorithm can suppress this overhead to $\mathcal{O}(\kappa\, \text{polylog}(\kappa/\epsilon))$ using variable-time amplitude amplification (VTAA) \cite{childs2017quantum}.}
\label{tab:qlss_scaling}
\end{table}

\subsubsection{Childs-Kothari-Somma (CKS) Algorithm}
\begin{figure}[htbp]
    \centering
    \includegraphics[width=\columnwidth]{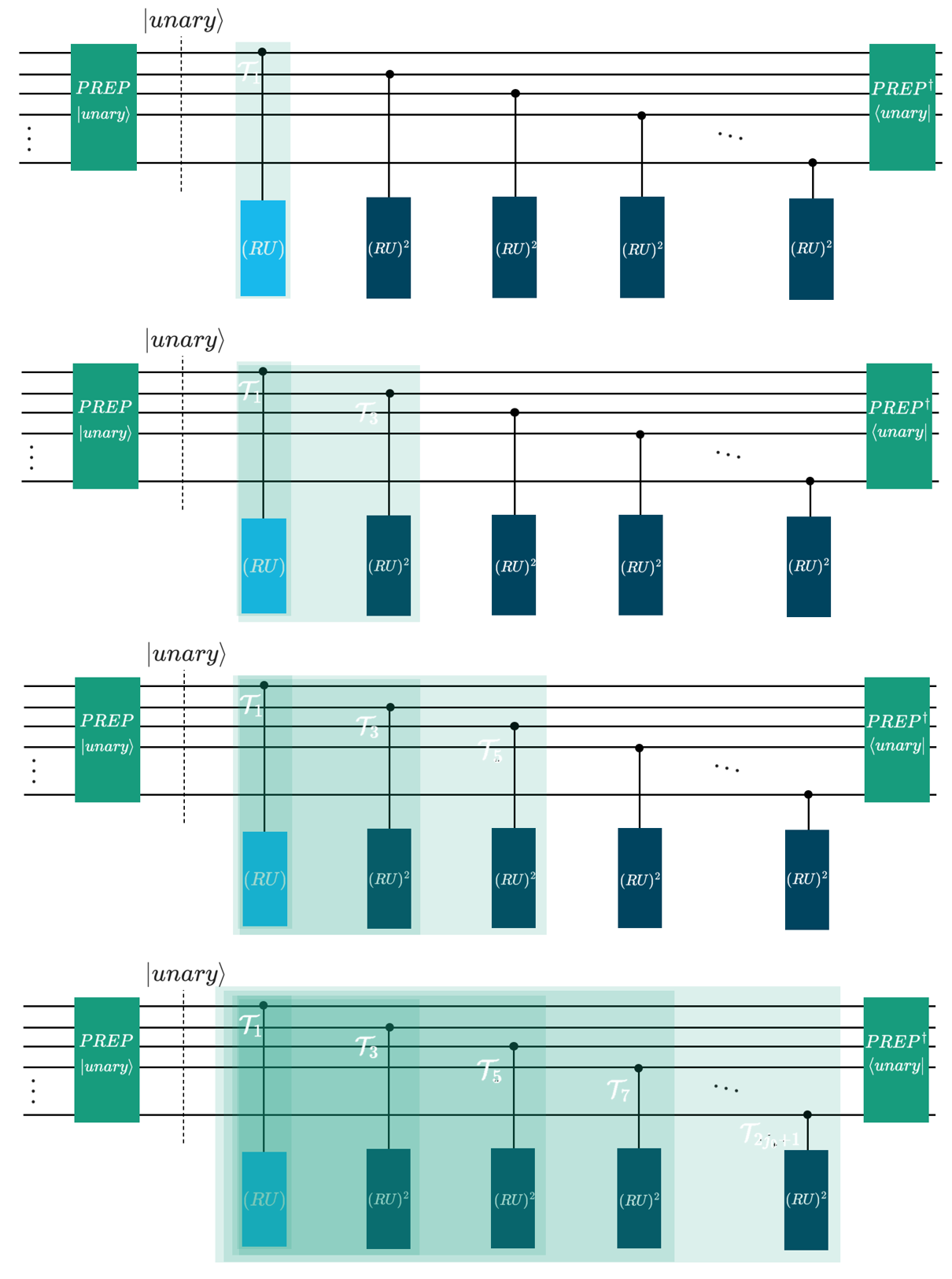}
    \caption{Visual schematic illustrating the efficient block-encoding of the $k$-th Chebyshev polynomial of the first kind $T_k(\hat{A})$ utilizing a unary control state. From top to bottom, conditioning on specific portions of the $\ket{unary}$ state sequentially triggers the appropriate powers of the walk operator $\hat{\mathcal{W}}$ to block-encode $T_k(\hat{A})$.}
    \label{fig:unary}
\end{figure}

\begin{figure}[htbp]
    \centering
    \includegraphics[width=\columnwidth]{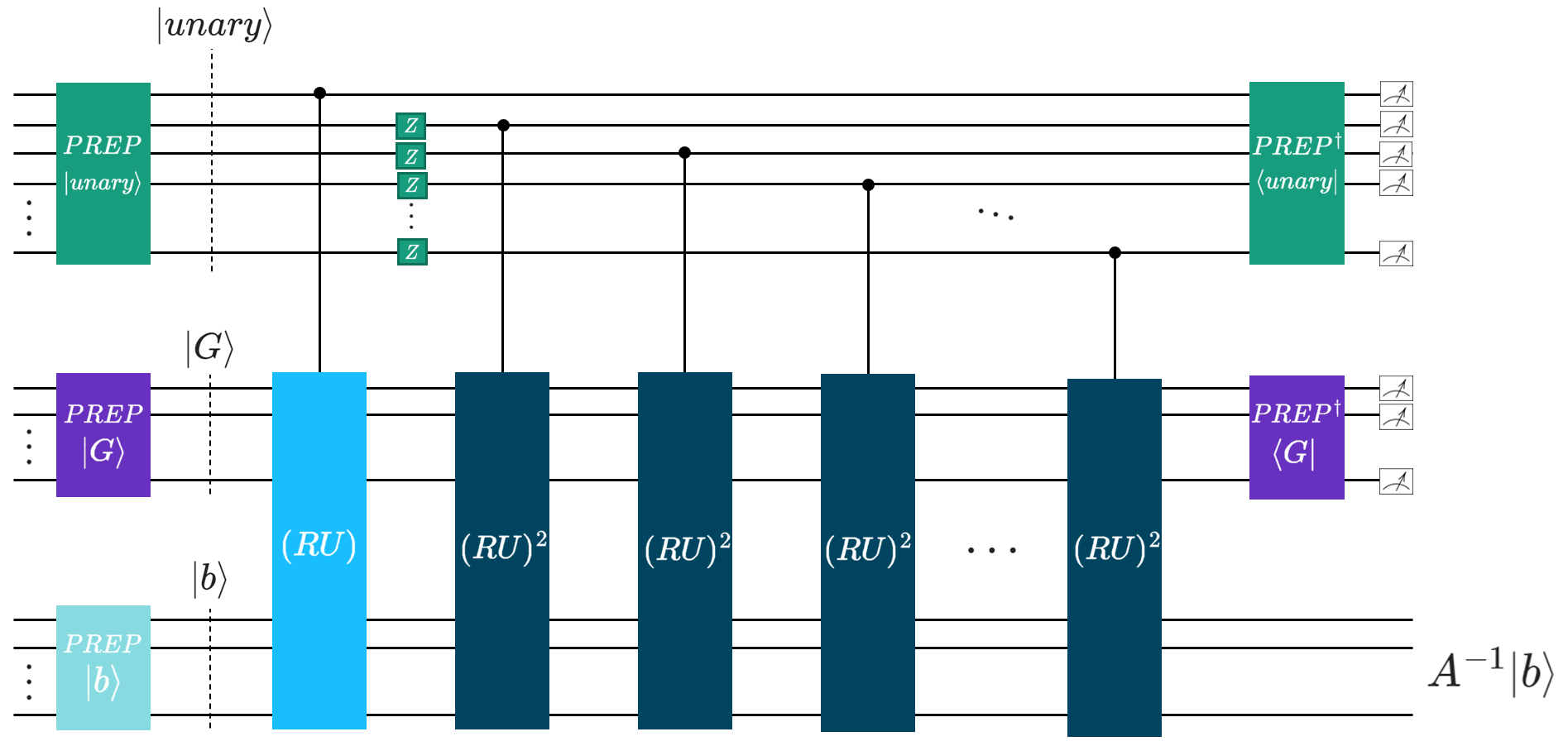}
    %\caption{Visual schematics of the Childs-Kothari-Somma (CKS) algorithm constructed by the \texttt{CKS(BE, eps, kappa)} function.}
    \caption{Visual schematic of the CKS algorithm implementation. The architecture leverages a Linear Combination of Unitaries (LCU) framework to evaluate the polynomial approximation of the matrix inverse.}
    \label{fig:CKS}
\end{figure}

The CKS algorithm \cite{childs2017quantum} achieves an exponentially improved dependence on precision compared to the HHL algorithm. It approximates the inverse by constructing a Linear Combination of Unitaries (LCU) corresponding to a weighted sum of Chebyshev polynomials, $\sum_k c_k T_k(\hat{A})$. In the Qrisp implementation (Figure~\ref{fig:CKS}), this is efficiently realized using an ancillary unary encoding. By encoding the coefficients $c_k$ into the amplitudes of a $\ket{unary}$ superposition state, individual qubits sequentially trigger powers of the qubitization walk operator $\hat{\mathcal{W}}(\hat{A})$. For instance, the basis state $\ket{1100}$ triggers $\hat{\mathcal{W}}$ and $\hat{\mathcal{W}}^2$, accumulating to $\hat{\mathcal{W}}^3$, which mathematically corresponds to the Chebyshev term $T_3(\hat{A})$ (Figure~\ref{fig:unary}). Operating in superposition evaluates the full linear combination simultaneously, requiring only $\mathcal{O}(d)$ applications of the base block-encoding for a degree-$d$ polynomial. The following example (Listing~\ref{lst:CKS}) demonstrates its execution on a well-conditioned $4 \times 4$ system.

\begin{lstlisting}[caption={CKS Linear Solver Implementation}, label={lst:CKS}]
import numpy as np
from qrisp import *
from qrisp.block_encodings import BlockEncoding
from qrisp.algorithms.cks import CKS

# Define a well-conditioned 4x4 system
A = np.array([[ 0.73,  0.15, -0.15, -0.04],
              [ 0.15,  0.69, -0.05, -0.01],
              [-0.15, -0.05,  0.77, -0.03],
              [-0.04, -0.01, -0.03,  0.59]])
b = np.array([0, 1, 1, 1])
kappa = np.linalg.cond(A)

BA = BlockEncoding.from_array(A)

# Instantiates and prepares the state b
def prep_b():
    qv = QuantumFloat(2)
    prepare(qv, b)
    return qv

@terminal_sampling
def run_cks():
    # CKS automatically calculates Chebyshev 
    # coefficients and builds the LCU circuit
    BA_CKS = CKS(BA, eps=0.01, kappa=kappa)
    return BA_CKS.apply_rus(prep_b)()

res_cks = run_cks()
amps_cks = np.sqrt([res_cks.get(i, 0) for i in range(len(b))])
print(amps_cks)
# [0.02737316 0.55866412 0.52852854 0.63859431]
\end{lstlisting}

\subsubsection{Dalzell's (Near-) Optimal Quantum Linear Solver}
The Dalzell solver \cite{dalzell2024shortcut} achieves asymptotically optimal scaling with respect to the condition number, significantly reducing the constant prefactors compared to previous state-of-the-art methods. Instead of relying on difficult-to-analyze techniques like variable-time amplitude amplification or adiabatic path-following, the algorithm simplifies the inversion by augmenting the linear system and using straightforward applications of kernel reflection via QSVT. Notably, this implementation requires an estimate of the solution norm $t \approx \|x\|$ as an input parameter. 

The following example (Listing~\ref{lst:dalzell}) illustrates how to invoke this solver\footref{gqsvt_branch}, reflecting the syntax developed in the experimental GQSVT branch of Eclipse Qrisp.

\begin{lstlisting}[caption={Dalzell Optimal QLSS Implementation}, label={lst:dalzell}]
import numpy as np
from qrisp import *
from qrisp.block_encodings import BlockEncoding
# Import from the GQSVT branch
from qrisp.algorithms.gqsp import dalzell_inversion

# Same 4x4 matrix A and vector b as defined above
BA = BlockEncoding.from_array(A)

# Prepares the state b on an existing variable
def prep_b(qv):
    prepare(qv, b)

# The Dalzell solver requires an estimate of the norm of the solution vector x
# For demonstration, we compute it classically:
x_exact = np.linalg.solve(A, b)
t_est = np.linalg.norm(x_exact) / np.linalg.norm(b)

@terminal_sampling
def run_dalzell():
    # The solver augments the linear system and applies kernel reflection via GQSVT
    BA_dalzell = dalzell_inversion(BA, 
                                   prep_b=prep_b,
                                   t=t_est,
                                   eps=0.01, 
                                   kappa=kappa)

    # Apply to the state |0>                               
    return BA_dalzell.apply_rus(lambda : QuantumFloat(2))()

res_dalzell = run_dalzell()
amps_dalzell = np.sqrt([res_dalzell.get(i, 0) for i in range(len(b))])
print(amps_dalzell)
# [0.02309097 0.56340042 0.52714249 0.63574175]
\end{lstlisting}

\section{Conclusion and Future Work}

In this work, we have introduced the expansion of the Eclipse Qrisp framework to introduce block-encodings as a high-level programming abstraction. The \texttt{BlockEncoding} interface lowers the bar for access to advanced quantum algorithmic techniques. By abstracting away manual ancilla allocation, subnormalization tracking, and circuit synthesis for block-encoded operators, it allows researchers to focus on high-level algorithm design.

However, the roadmap for Qrisp is not a closed internal trajectory, but an invitation. In the spirit of \textbf{open source, open development, and open research}, the framework is designed to serve as a collaborative ecosystem. While the following functionalities represent high-impact directions for quantum numerical linear algebra, they are intended as community-driven objectives where any researcher or developer is encouraged to contribute.

Future development will focus on three primary research directions:

\subsection{Community-Driven Toolbox: adding custom block-encodings as methods of the \texttt{BlockEncoding} class}
To provide a comprehensive ``toolbox'' for custom block-encodings, one might want to implement further (efficient for their respective purpose) block-encoding techniques. This includes the implementation of oracle-based methods such as FABLE \cite{camps2022fable} and its spiritual successors, S-FABLE and LS-FABLE \cite{kuklinski2024s}. One could also implement FOCQS-LCU \cite{della2025efficient} \cite{nibbi2026practical}. For sparse and structured matrices, one could implement dictionary-based encodings \cite{yang2025dictionary}, stabilizer-formalism-based block-encodings for Pauli strings \cite{schillo2026block}, and explicit block-encodings for discrete Laplacians with mixed boundary conditions \cite{laplacian2026}.

\subsection{Quantum Numerical Linear Algebra and Matrix Equations}
Another path for extending the \texttt{BlockEncoding} class is the one to solve complex linear matrix equations. One could implement solvers for the Sylvester equation ($AX + XB = C$) \cite{somma2025quantum} by utilizing vectorization techniques described in \cite{dong2025products}. This direction can also include:
\begin{itemize}
    \item \textbf{Quantum Oracle Sketching} for processing massive classical data\cite{zhao2026exponential};
    \item \textbf{Quantum Support Vector Machines} for big data classification \cite{rebentrost2014quantum};
    \item \textbf{Quantum Regularized Least Squares} for robust data fitting \cite{chakraborty2023quantum};
    \item \textbf{Preconditioned Block Encodings} to improve the condition number of linear systems \cite{lapworth2025preconditioned};
    \item Efficient algorithms for \textbf{anisotropic diffusion and convection equations} using vector norm scaling \cite{zylberman2025quantum}.
\end{itemize}

\subsection{Quantum Chemistry and Many-Body Dynamics}
We are dedicated to enabling simulation of complex physical systems through resource-efficient block-encodings. This includes:
\begin{itemize}
    \item \textbf{Counderdiabatic driving} By efficiently block-encoding nested commutators, one could also apply it to Counterdiabatic driving \cite{finvzgar2025counterdiabatic}, \cite{chen2022efficient};
    \item \textbf{Green's Functions and Dipole Operators:} Implementing block-encodings for Green's functions useful for fermionic systems \cite{loaiza2026quantum};
    \item \textbf{Electronic Structure:} Targeting Double Factorization \cite{von2021quantum} and Tensor Hypercontraction (THC) \cite{lee2021even};
    \item \textbf{Complex Dynamics:} Expanding support for Pre-Born-Oppenheimer dynamics \cite{pocrnic2026efficient}, non-adiabatic dynamics at metallic surfaces \cite{lang2026quantum}, and vibronic dynamics for molecular design \cite{motlagh2025quantum}.
\end{itemize}

Ultimately, the \texttt{BlockEncoding} interface in Eclipse Qrisp establishes a vital separation of concerns. While constructing efficient, specialized block-encodings inherently demands deep quantum algorithmic expertise, our framework ensures that this rigorous development effort translates directly into broad scientific utility. By encapsulating these complex mathematical constructs into high-level programming abstractions, newly implemented encodings become readily deployable for the wider community. Consequently, expanding the native library of encodings is a highly impactful avenue for future work, as it directly empowers domain researchers to seamlessly utilize advanced quantum numerical linear algebra subroutines without managing the underlying quantum mechanics.

The authors take liberty to provide food for thought with such an absurd claim: ``The future of programming quantum numerical linear algebra is here and it's Qrisp. Eclipse Qrisp''.

\section*{Acknowledgments}
This work was funded by the Federal Ministry of Research, Technology and Space (German: Bundesministerium für Forschung, Technologie und Raumfahrt; abbreviated BMFTR) under the project grant FullStaQD (grant agreement number 01MQ25001A), and by the European Union under project grant SecQdevOps (grant agreement number 101225776). The authors are responsible for the content of this publication.

\smallskip

The authors would like to thank Marek Gluza for drawing our attention to the quantum Lanczos paper, which was instrumental in our conceptualization of block encodings, qubitization, and block-encoded Chebyshev polynomials. We also recognize Dong An, whose seminar on Linear Combinations of Hamiltonian Simulations (LCHS) served as the primary catalyst for the authors’ venture into implementing block-encodings. We are grateful to Nathan Wiebe for his insightful online lectures and for highlighting the use of unary encoding in his recorded QSim 2025 talk. We also thank Lin Lin for his in-depth lecture notes on block encodings and Quantum Signal Processing (QSP), and Alexander Dalzell for his seminar on his (near-) optimal linear solver, which provided the foundation for our implementation. Special thanks are due to Raphael Seidel and Pietropaolo Frisoni for implementing the methods invoked in \texttt{.resources()} and for their thorough review of the \texttt{BlockEncoding} implementation. Furthermore, we thank Danial Motlagh, Andrew Childs, Robin Kothari, Lorenzo Laneve, Christoph Sünderhauf, Jiaqi Leng, Jernej Rudi Finžgar,  Antigoni Georgiadou, Murali Gopalakrishnan Meena, Nikolay Tcholtchev, and Sebastian Bock for their insightful correspondences and discussions.

M.\ P.\ would like to additionally thank Nathan Wiebe for a fruitful brainstorming discussion at QCTiP 2025 in Berlin, specifically brainstorming about ``how cool would it be to provide block-encodings as programming abstractions [... in Eclipse Qrisp]''. Regarding the software's design and implementation, M.\ P.\ would like to thank Matija Pretnar for discussions on centering the implementations within a unified class structure, and Martina Nibbi for valuable exchanges regarding FOQCS-LCU. Appreciation should also go towards Leon Alexander Rullkötter, Niclas Schillo, and Andreas Sturm for discussing the integration of their respective approaches as core methods of the \texttt{BlockEncoding} class. Furthermore, M.\ P.\ thanks Jure Leskovec for an insightful conversation on the abstractions and entry points necessary for classical developers to engage with the field, while also taking hats off to Matteo Antonio Inajetovic and Diego Polimeni for their early support and feedback on these ideas. Finally, the authors thank Jens Palsberg for a plethora of insightful correspondences following a seminar presented to his research group.

The authors acknowledge the use of Gemini 3.1 Pro (Google DeepMind), as well as Claude Haiku 4.5 via GitHub Copilot, for assistance in drafting and refining the manuscript from the existing docstrings and other documentation \cite{qrisp_docs}. The authors take full responsibility for the content.

\section*{Code availability}
\textit{Eclipse Qrisp} is an open-source Python framework for high-level programming of quantum computers.
The source code is available in \href{https://github.com/eclipse-qrisp/Qrisp}{https://github.com/eclipse-qrisp/Qrisp}. Specialized features, including the GQSVT-based linear solvers, can be found in the repository's \texttt{GQSVT} branch: \href{https://github.com/eclipse-qrisp/Qrisp/tree/GQSVT}{https://github.com/eclipse-qrisp/Qrisp/tree/GQSVT}.

\nocite{*}

\bibliographystyle{IEEEtran}
\bibliography{sources}
\end{document}